\documentclass[useAMS,usenatbib]{mnras}

\usepackage{graphicx}
\usepackage{fixltx2e}
\usepackage{amssymb}
\title[MHS modelling of stellar coronae]{Magnetohydrostatic modelling of stellar coronae}
\author[D. MacTaggart et al.]{D. MacTaggart$^{1}$\thanks{E-mail:
david.mactaggart@glasgow.ac.uk}, S.G. Gregory$^{2}$, T.
Neukirch$^{3}$ and J.-F. Donati$^{4}$\\
$^{1}$School of Mathematics and Statistics, University of Glasgow, Glasgow, G12 8QW, UK\\
$^{2}$SUPA, School of Physics and Astronomy, University of St Andrews, St Andrews, KY16 9SS, UK\\
$^{3}$School of Mathematics and Statistics, University of St Andrews, St Andrews KY16 9SS, UK\\
$^{4}$LATT-UMR 5572, CNRS \& Univ. Paul Sabatier, 14 Av. E. Berlin, F-31400 Toulouse, France}
\begin{document}

\date{Accepted 1988 December 15. Received 1988 December 14; in original form 1988 October 11}

\pagerange{\pageref{firstpage}--\pageref{lastpage}} \pubyear{2002}

\maketitle

\label{firstpage}

\begin{abstract}
We introduce to the stellar physics community a method of modelling stellar coronae that can be considered to be an extension of the potential field. In this approach, the magnetic field is coupled to the background atmosphere. The model is magnetohydrostatic (MHS) and is a balance between the Lorentz force, the pressure gradient and gravity. Analytical solutions are possible and we consider a particular class of equilibria in this paper. The model contains two free parameters and the effects of these on both the geometry and topology of the coronal magnetic field are investigated. A demonstration of the approach is given using a magnetogram derived from Zeeman-Doppler imaging of the 0.75$\,{\rm M}_{\odot}$ M-dwarf star GJ~182.
\end{abstract}


\section{Introduction}
The atmospheres of cool stars are governed by their magnetic fields. Phenomena ranging from flares to coronal mass ejections (CMEs) to stellar winds are all intimately related to the atmospheric magnetic field. The best case study we have for a stellar magnetic field is the Sun. The solar magnetic field has been studied in detail over many scales and cycles. We cannot, however, observe the atmospheric (coronal) magnetic field directly and so must rely upon theoretical models. As input to these models, information about the magnetic field at the Sun's surface (photosphere) can be determined through the Zeeman effect \citep{priest82}. Maps of the line-of-sight magnetic field (magnetograms) have been available for many years. More recently, vector magnetograms of the Sun have also been incorporated into coronal models \citep[e.g.][]{wiegelmann12, regnier13}.

On stars other than the Sun, we now have insight into magnetic field structures. We can use Zeeman Doppler Imaging (ZDI) to determine stellar magnetograms \citep[e.g.][]{donati08}. Such magnetograms convey only the large-scale structure of a star's magnetic field. Fields on the active region scale and smaller are not observable currently. Much important information about the stellar enviroment, however, can be gained from a knowledge of the large-scale structure of the global magnetic field. Two approaches have been used mainly to study the magnetic coronae of stars (henceforth `stars' refers to stars other than the Sun). The first is a potential field extrapolation \citep[e.g.][]{jardine02,lang12,gregory11}. The potential field approach is useful as it gives a unique solution that minimizes the magnetic energy for a given set of boundary conditions. However, as there is no current ($\mathbf{j}=\mathbf{0}$) in a potential field, the magnetic field in the corona does not contain twist, which is considered important for the onset of different forms of eruptive behaviour \citep[e.g.][]{hood11}. Also, the Lorentz force completely decouples from the non-magnetic terms in the magnetohydrodynamic (MHD) momentum equation. This decoupling means that certain assumptions need to be made when finding proxies for quantities such as X-ray emission \citep[e.g.][]{jardine08}. The second approach is to perform a full MHD simulation of the global corona \citep[e.g.][]{riley06,cohen10,romanova10,vidotto11},   where it is possible to capture the large-scale dynamics of the corona. However, this approach is computationally very expensive and often requires super-computing resources.

In this paper, we introduce a technique that can be thought of as an intermediary step between the potential field model and full MHD simulations, as mentioned above. We present a magnetohydrostatic (MHS) solution which couples the coronal density, pressure and magnetic field together. Perturbations in the hydrostatic density and pressure, due to the magnetic field, can be calculated self-consistently. As the solution is static, like the potential field model, it is very simple to implement. The model allows us to calculate, with relative ease, a twisted stellar coronal field where magnetic forces balance hydrostatic forces. The approach presented in this paper has been applied to the solar corona where large-scale features have been captured successfully \citep{bogdan86,neukirch95,zhao00,ruan08}. Our approach will be particularly suited to stellar applications, where magnetograms only convey the large-scale structure of the magnetic field. This type of model can capture the large-scale field structure (as evidenced by the solar applications mentioned above) and improve upon the potential field extrapolations with the inclusion of current density and the coupling of the magnetic field to the background stellar atmosphere. 

 In the following section we give a detailed derivation of the coronal magnetic field and the associated density and pressure perturbations (correcting typographical errors that appear in previous works). We follow this section with an analysis of how the resulting free parameters affect the geometry and topology of the model coronal magnetic field. After the  analysis, an application of the model is given where a magnetic map of the M-dwarf GJ~182, published by \cite{donati08},  is used as input. The paper ends with a summary.

\section{Atmosphere model}
\subsection{Magnetic field}
The MHS equations that describe the balance between the Lorentz and hydrostatic forces can be written as
\begin{equation}\label{momentum}
\mathbf{j}\times\mathbf{B}-\nabla p-\rho\mathbf{g} = \mathbf{0}, \\
\end{equation}
\begin{equation}\label{ampere}
\nabla\times\mathbf{B} = \mu_0\mathbf{j}, \\
\end{equation}
\begin{equation}\label{divB}
\nabla\cdot\mathbf{B} = 0.
\end{equation}
Here, $\mathbf{B}$ is the magnetic induction (referred to as the magnetic field), $\mathbf{j}$ is the curent density, $p$ is the plasma presure, $\rho$ is the plasma density and $\mathbf{g}$ is gravity. A discussion of appropriate boundary conditions will be left until later. Rather than proceeding with equations (\ref{momentum}) to (\ref{divB}) in dimensional form, we write them in dimensionless form by introducing the variables
\begin{eqnarray}\label{nondim_p}
&&\mathbf{r} = r_0\mathbf{r}^*, \quad\mathbf{B} = B_0\mathbf{B}^*,\quad \mathbf{g}=g_0\mathbf{g}^*, \nonumber\\ 
&& p = (B_0/\mu_0)p^*,  \quad \rho = B_0^2/(\mu_0r_0g_0)\rho^*,  \\
&&\mathbf{j} = B_0/(\mu_0r_0)\mathbf{j}^*. \nonumber
\end{eqnarray}
The distance vector is labelled $\mathbf{r}$ in anticipation of the use of spherical coordinates for stellar applications. Dropping the asterisks, the dimensionless MHS equations are
\begin{equation}\label{mom_nd}
\mathbf{j}\times\mathbf{B}-\nabla p-\rho\nabla\psi = \mathbf{0}, \\
\end{equation}
\begin{equation}\label{ampere_nd}
\nabla\times\mathbf{B} = \mathbf{j}, \\
\end{equation}
\begin{equation}\label{divB_nd}
\nabla\cdot\mathbf{B} = 0.
\end{equation}
 All further expressions will be in dimensionless form unless stated otherwise. Here we have written gravity in terms of its potential $\psi$, viz.
\begin{equation}
\psi = -\frac{1}{r},
\end{equation}
To make progress, we model the current density as the composition of two distinct terms:
\begin{equation}\label{current}
\mathbf{j} = \alpha\mathbf{B} + \nabla\times(F\mathbf{r}).
\end{equation}
 This choice of the current density enables us to model the magnetic field through a range of plasma $\beta~ (=2\mu_0p/B_0^2)$. Since equation (\ref{current}) is linear, we can make analytical progress and take a practical intermediate step between potential field extrapolations and expensive global MHD simulations. $\alpha$ is a constant parameter and the first term on the RHS of equation (\ref{current}) represents the current density of a linear force-free field. The second term on the RHS represents the current density that produces a Lorentz force that can balance the effects of the pressure gradient and gravity. $F$ is a free function which will be specified shortly. The approach of using this type of term was introduced by \cite{low82} and has been used in several subsequent studies \citep[e.g.][]{low91,gibson96,neukirch99,petrie00}. \cite{jardine13} also present a non-potential magnetic field model, in the stellar physics context, with a term similar to the second term on the RHS of equation (\ref{current}). However, they do not consider MHS equilibria. The combination of the two terms in equation (\ref{current}) results in components of current density that are both parallel and perpendicular to the magnetic field. Therefore, the total field-aligned current density, if written in the form $\bar{\alpha}\mathbf{B}$, has a non-uniform $\bar{\alpha}$.  Although the field in our model can be (depending on the choice of parameters discussed below) more complicated than a linear force-free field, our model has the advantage of taking the same input from magnetograms as linear force-free models. As potential field models are a particular case of linear force-free models $(\alpha =0)$,  potential models can be updated to the MHS model easily.

The following derivation of the model is based on \cite{bogdan86} and \cite{neukirch95}. An alternative method exploiting the toroidal and poloidal components of the field can also be used \citep{neukirch99}. We shall utilize the toroidal-poloidal approach later when studying the topology of the field.

Inserting equation (\ref{current}) into Amp\`{e}re's law, equation (\ref{ampere}), gives, after some manipulation,
\begin{equation}\label{rdot}
\mathbf{r}\cdot(\nabla\times\nabla\times\mathbf{B}) = \alpha^2\mathbf{r}\cdot\mathbf{B}+\mathbf{r}\cdot\nabla\times(\nabla F\times\mathbf{r}),
\end{equation}
where $\mathbf{r}$ is the radial vector. Further, we choose the form of $F$ to be
\begin{equation}\label{Fdefine}
F=\xi(r)\mathbf{r}\cdot\mathbf{B},
\end{equation}
where $\xi(r)$ is a free function.  The particular form of $\xi(r)$ will be chosen later to allow for an analytical solution. At this point it is useful to introduce the angular momentum operator
\begin{equation}\label{L1}
\mathbf{L} \equiv -i\mathbf{r}\times\nabla.
\end{equation}
This operator is common in quantum mechanics and a description of its properties can be found in \cite{jackson75}. By simple manipulation, it can be shown that
\begin{equation}\label{tmp1}
\mathbf{r}\cdot(\nabla\times\nabla\times\mathbf{B}) = -\nabla^2(\mathbf{r}\cdot\mathbf{B}).
\end{equation}
 Combining equations (\ref{rdot}) and  (\ref{tmp1}), and making use of  equations (\ref{Fdefine}) and (\ref{L1}), gives
\begin{equation}\label{vec_schrod}
\nabla^2(\mathbf{r}\cdot\mathbf{B}) + {\xi}(r)\mathbf{L}^2(\mathbf{r}\cdot\mathbf{B})+ {\alpha}^2(\mathbf{r}\cdot\mathbf{B}) = 0.
\end{equation}
To proceed, we expand $\mathbf{r}\cdot\mathbf{B}$ in spherical harmonics,
\begin{equation}\label{rdotB}
\mathbf{r}\cdot\mathbf{B} = \sum_{l=1}^{\infty}\sum_{m=-l}^ll(l+1)Q_{lm}Y^m_l(\theta,\phi),
\end{equation}
where
\begin{equation}
Q_{lm} = \sum_{j=1}^2 A^{(j)}_{lm}u^{(j)}_{l}(r).
\end{equation}
$A^{(j)}_{lm}$ are constant (complex) coefficients. To determine the $u^{(j)}_{l}(r)$, inserting equation (\ref{rdotB}) into equation (\ref{vec_schrod}) gives
\begin{equation}\label{g}
\left[\frac{{\rm d}^2 }{{\rm d}r^2} -l(l+1)\left(\frac{1}{r^2}-\xi(r)\right) + {\alpha}^2\right]g_l^{(j)}=0.
\end{equation}
where $g_l^{(j)} = ru^{(j)}_{l}$. In deriving equation (\ref{g}), use has been made of the property
\begin{equation}\label{L}
\mathbf{L}^2Y^m_l(\theta,\phi) = l(l+1)Y^m_l(\theta,\phi).
\end{equation}
To solve equation (\ref{g}) analytically, a particular form must be chosen for $\xi(r)$.  Notice that equation (\ref{g}) has the form of a 1D Schr\"{o}dinger equation with $\xi(r)$ playing the role of the potential \citep[e.g.][]{rae02}. This analogy gives us a starting point for what form $\xi(r)$ can take  in order to yield analytical solutions. Several possible analytical solutions are discussed in \cite{neukirch95}.  Following \cite{ruan08}, we adopt
\begin{equation}\label{xi}
\xi(r) = \frac{1}{r^2}-\frac{1}{(r+a)^2},
\end{equation}
where $a$ is a free parameter. This particular form represents the contribution of the non-force-free part of the current density decaying with height, as in the low solar corona \citep[e.g.][]{gary01}. By considering the following variable transformation:
\begin{equation}
s = r+a, \quad f_l^{(j)} =   \frac{g_l^{(j)}}{s^{1/2}},
\end{equation}
equation (\ref{g}) transforms into Bessel's equation
\begin{equation}
\left[\frac{{\rm d}^2 }{{\rm d}s^2} + \frac{1}{s}\frac{\rm d}{{\rm d}s} + \alpha^2 - \frac{(l+1/2)^2}{s^2}\right]f^{(j)}_l = 0.
\end{equation}
Converting back to the original variables, the solutions can be written as
\begin{eqnarray}
u^{(1)}_{l} &=& \frac{\sqrt{r+a}}{r}J_{l+1/2}(\alpha(r+a)), \label{u1}\\
 u^{(2)}_{l} &=& \frac{\sqrt{r+a}}{r}N_{l+1/2}(\alpha(r+a)) \label{u2}, 
\end{eqnarray}
where $J_{l+1/2}$ and $N_{l+1/2}$ are Bessel functions of the first and second kind respectively.

Before writing down an expression for the magnetic field, we require the $A^{(j)}_{lm}$ which are determined using the boundary conditions and an orthogonality constraint. As mentioned earlier, the solutions of the MHS equations (\ref{momentum})-(\ref{divB}) require boundary conditions. In this problem, the required boundary conditions are the radial components of the field at the stellar surface and a source surface. At the stellar surface, the radial magnetic field component is provided by a magnetogram. The modeller is free to select an appropriate form for the source surface and choose the distance of the source surface from the stellar surface. In solar applications, in-situ satellite observations can constrain the form of the source surface. No such constraints exist for stellar coronae, although X-ray observations can be used to provide loose constraints \citep[e.g.][]{hussain07}. For MHS models, the effects of outflows (e.g. stellar winds) are incorporated into the form of the source surface. At the stellar surface $r=1$ (we have taken $r_0=R_{\ast}$) the radial component of the field, $B^1_r$, can be written as
\begin{equation}
B_r^1(\theta,\phi)=\sum^{\infty}_{l=1}\sum^l_{m=-l}l(l+1)Q^1_{lm}Y^m_l(\theta,\phi),
\end{equation}
where
\begin{eqnarray}
Q^1_{lm} &=& A^{(1)}_{lm}\sqrt{1+a}J_{l+1/2}(\alpha(1+a)) \nonumber \\
&+& A^{(2)}_{lm}\sqrt{1+a}N_{l+1/2}(\alpha(1+a)).
\end{eqnarray}
Note that if the magnetogram comes from observations, it is important to ensure that the magnetic flux is balanced. Using the orthogonality condition
\begin{equation}
\int_{\theta=0}^{\pi}\int_{\phi=0}^{2\pi}Y^m_l\bar{Y}^{m'}_{l'}\sin\theta\,{\rm d}\theta{\rm d}\phi = \delta_{ll'}\delta_{mm'},
\end{equation}
 we find
\begin{equation}\label{q1}
Q^1_{lm} = \frac{1}{l(l+1)}\int_{\theta=0}^{\pi}\int_{\phi=0}^{2\pi}B_r^1(\theta,\phi)\bar{Y}^m_l\sin\theta\,{\rm d}\theta{\rm d}\phi .
\end{equation}
Here, the overbar represents the complex conjugate. Equation (\ref{q1}) is a linear equation for $A^{(1)}_{lm}$ and $A^{(2)}_{lm}$. Applying the same procedure at the source surface, or imposing another constraint, results in another linear equation and this system can be solved to find the coefficients $A_{lm}^{(j)}$. 

To find a closed-form solution for the magnetic field, we take the dot product of $\mathbf{r}$ and equation (\ref{ampere_nd}) and apply the angular momentum operator,
\begin{equation}\label{lb}
\mathbf{L}\cdot\mathbf{B} = -i\alpha\mathbf{r}\cdot\mathbf{B}.
\end{equation}
From the definition of $\mathbf{L}$ in equation (\ref{L1}), equation (\ref{lb}) contains only derivatives of $B_{\theta}$ and $B_{\phi}$. Hence,  we can write
\begin{equation}\label{LB}
\mathbf{L}\cdot\mathbf{B}_T = -i\alpha\sum_{l=1}^{\infty}\sum_{m=-l}^l\sum_{j=1}^2l(l+1)A^{(j)}_{lm}u^{(r)}_l(r)Y^m_l(\theta,\phi),
\end{equation}
where $\mathbf{B}_T = B_{\theta}\mathbf{e}_{\theta}+B_{\phi}\mathbf{e}_{\phi}$. We require a series expansion of $\mathbf{B}_T$ to match the RHS of equation (\ref{LB}). Following Jackson (1975) and Neukirch (1995), we find
\begin{equation}\label{BT}
\mathbf{B}_T = \sum_{l=1}^{\infty}\sum_{m=-l}^l v_{lm}(r)\mathbf{L}Y^m_l(\theta,\phi)+w_{lm}\nabla Y^m_l(\theta,\phi).
\end{equation}
The unknown functions $v_{lm}(r)$ and $w_{lm}(r)$ can be found by two simple considerations. For brevity, we shall consider a single order for $l$ and $m$. Inserting equation (\ref{BT}) into equation (\ref{LB}) and making use of equation (\ref{L}) results in
\begin{equation}
v_{lm}(r) =  -i\alpha\sum_{j=1}^2{A^{(j)}_{lm}}u^{(j)}_l(r).
\end{equation}
On the application of equation (\ref{divB_nd}), and some simple rearrangement, we find
\begin{equation}
w_{lm} = \sum_{j=1}^2A^{(j)}_{lm}\frac{\rm d}{{\rm d} r}\left(ru^{(j)}_l(r)\right).
\end{equation}
The expression for the global coronal magnetic field is then given by
\begin{eqnarray}\label{globalB}
\mathbf{B} &=& \sum_{l=1}^{\infty}\sum_{m=-l}^l\sum_{j=1}^2A^{(j)}_{lm}\left[l(l+1)u_l^{(j)}\frac{\mathbf{r}}{r^2}Y^m_l(\theta,\phi)\right. \nonumber  \\
&-&\left. {i{\alpha}}u_l^{(j)}\mathbf{L}Y^m_l(\theta,\phi)+\frac{\rm d}{{\rm d}r}\left(ru_l^{(j)}\right)\nabla Y^m_l(\theta,\phi)\right]
\end{eqnarray}
The spherical components of equation (\ref{globalB}) are listed in Appendix \ref{appendix}.

\subsection{Non-magnetic variables}
We now consider what forms the pressure and density take in the MHS model. Inserting the model current density, given by equation (\ref{current}), into the force balance equation (\ref{mom_nd}) gives, after some rearrangement,
\begin{equation}\label{insertion}
\nabla p + \rho\nabla\psi+(\mathbf{B}\cdot\mathbf{r})\nabla F - (\mathbf{B}\cdot\nabla F)\mathbf{r} = \mathbf{0}.
\end{equation}
For a non-trivial current density, we require $\nabla F$ and $\mathbf{r}$ to be linearly independent. Hence, equation (\ref{insertion}) implies that $p=p(\mathbf{r},F)$. Considering the component in the direction $\nabla F$, we obtain
\begin{equation}\label{pF}
\left.\frac{\partial p}{\partial F}\right|_{\mathbf{r}} = -\mathbf{B}\cdot\mathbf{r} = -\frac{1}{\xi(r)}F.
\end{equation}
Integrating equation (\ref{pF}) yields
\begin{equation}\label{pressure}
p = p_0(r) -\frac{1}{2}\xi(r)(\mathbf{r}\cdot\mathbf{B})^2.
\end{equation}
$p_0$ is the background pressure and is found from hydrostatic balance. The second term in equation (\ref{pressure}) is due to the non-force-free component in the current density. 

By considering the $\mathbf{r}$-component of equation (\ref{insertion}), the density is found to be
\begin{equation}\label{density}
\rho = \rho_0(r) + {r^2}\left(\frac{1}{2}\frac{{\rm d}\xi}{{\rm d} r}(\mathbf{r}\cdot\mathbf{B})^2+r\xi \mathbf{B}\cdot \nabla(\mathbf{r}\cdot\mathbf{B})\right).
\end{equation}
Here, $\rho_0(r)$ is the background hydrostatic density profile.

\section{Analysis}

In this section, we analyze the effects of the free parameters, $\alpha$ and $a$, on the geometry and topology of the coronal magnetic field. These parameters determine the form of the two model current density terms given in equation (\ref{current}).

\subsection{Free parameters and field geometry}
\begin{figure}
\includegraphics[width=\linewidth]{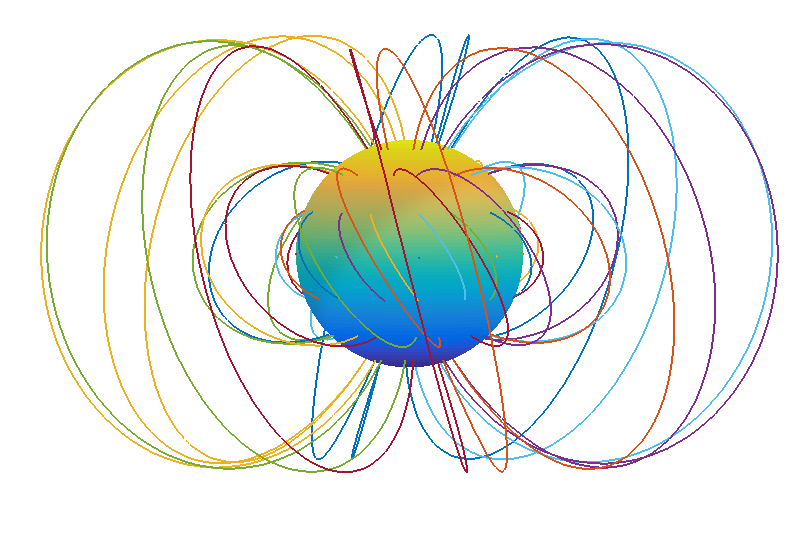}
\caption{ Field line plot of the dipolar field specified in equation (\ref{bcs}). The free parameters are $\alpha=0.5$ and $a=0.1$. Low magnetic arcades are aligned at different angles to those that rise to greater radial distances. This shearing of the field is due to a non-zero $\alpha$. The coloured stellar surface shows the non-dimensional $B_r$ component of the dipolar field in equation (\ref{bcs}). }
\label{twist}
\end{figure}

The main assumption of the MHS model is the form of the current density in equation (\ref{current}).  The first term that contributes to the current density, $\alpha \mathbf{B}$,  represents the linear force-free part of the current that is parallel to the magnetic field.  This term is well known for adding twist to the magnetic field.  Analogous to vorticity in a sheared flow, current density can manifest itself through a sheared magnetic field.  To demonstrate this effect, we shall consider a dipolar field with very simple boundary conditions. These are

\begin{equation}\label{bcs}
B_r(R_{\star},\theta,\phi)=\cos\theta, \,\, B(R_{\rm SS},\theta,\phi)=0,
\end{equation}
where $R_{\star}=1$ is the stellar surface and $R_{\rm SS}=3$ is the source surface. Figure \ref{twist} shows the coronal magnetic field with  boundary conditions (\ref{bcs}) and free parameters $\alpha=0.5$ and $a=0.1$. The effect of $\alpha\ne 0$ can be seen clearly. Moving from the stellar surface upwards, the alignment angle of the magnetic arcades changes. This shearing of the magnetic field is evidence of non-zero current.

\begin{figure}
\includegraphics[width=\linewidth]{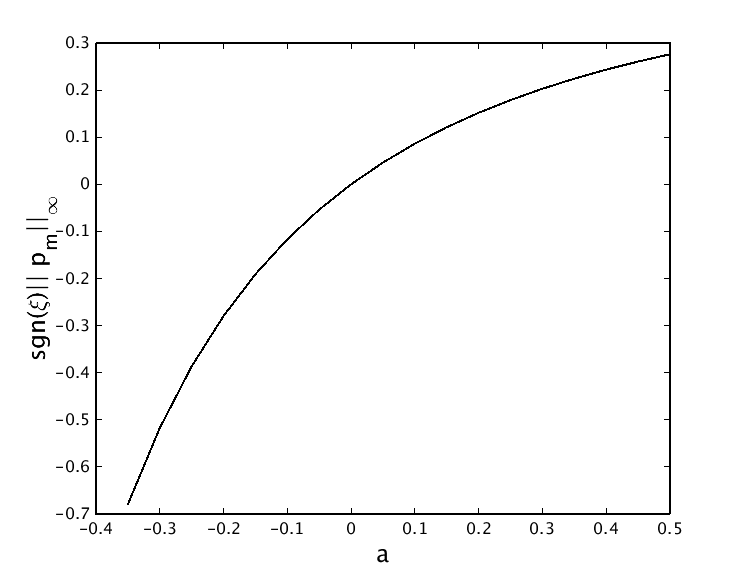}
\caption{ The variation of the signed maximum magnitude of the pressure perturbation $p_m$ as a function of $a$ for fixed $\alpha$=0.3. Values on the axes are non-dimensional as specified in equation (\ref{nondim_p}).}
\label{fig_pm}
\end{figure}

The second term in equation (\ref{current}) is perpendicular to gravity (the radial direction). It is this term that couples the magnetic field to the plasma pressure and density. The second term depends on both of the free parameters $\alpha$ and $a$. As mentioned above, increasing the magnitude of $\alpha$ increases the twist (positive and negative values determine the sense of the twist). The effect of changing $a$ requires more investigation. \cite{zhao00} perform several numerical experiments and show that increasing the value of $a$ causes magnetic arcades to expand. Physically, this expansion occurs if there is an increase in magnetic pressure. Consider equation (\ref{pressure}) re-written as
\begin{equation}\label{pp}
p+p_m = p+\frac{1}{2}\xi(r)(\mathbf{r}\cdot\mathbf{B})^2=p_0,
\end{equation}
where $p$ is the plasma pressure, $p_m$ is the magnetic pressure perturbation and $p_0$ is the background hydrostatic pressure of the stellar atmosphere. By considering the expression for the pressure perturbation, the sign of $p_m$ depends on the sign of $\xi(r)$. From equation (\ref{xi}), the sign depends, in turn, on the choice of $a$. $p_m$ also depends on $a$ through equations (\ref{u1}) and (\ref{u2}). To give an illustration of how the the pressure perturbation depends on $a$, we calculate the signed maximum perturbation, ${\rm sgn}(\xi)\|p_m\|_{\infty}$. This quantity can be found analytically as
\begin{equation}\label{maxmag}
{\rm sgn}(\xi)\|p_m\|_{\infty} = \frac{1}{2}\xi(1)=\frac{1}{2}\left(1-\frac{1}{(1+a)^2}\right),
\end{equation}
using the fact that the maximum magnitude of the pressure perturbation occurs at the stellar surface where $\|\mathbf{r}\cdot\mathbf{B}\|_{\infty}=1$. Figure \ref{fig_pm} shows a plot of equation (\ref{maxmag}).

For $a>0$, the values are positive, indicating an expansion of the magnetic field \citep[e.g.][]{zhao00}. When $a<0$, $p_m$ is also negative, resulting in compression of the magnetic field. What this analysis shows is that the choice of $\xi(r)$ is crucial for determining how the pressure perturbation will behave. The pressure perturbation, in turn, will influence {the inflation of the magnetic field}.

\subsection{Magnetic helicity}
Magnetic helicity is a topological invariant of ideal MHD \citep[e.g.][]{biskamp97} and can be interpreted as the average pairwise linkage of magnetic field lines \citep{arnold92}. The magnetic helicity is defined by
\begin{equation}\label{helicity}
H=\int_V\mathbf{B}\cdot\mathbf{A}\,{\rm d}V,
\end{equation}
where $\mathbf{A}=\nabla\times\mathbf{B}$ and $V$ is the volume of concern. This quantity is only gauge-invariant, however, if the magnetic field is closed, that is if $\mathbf{B}\cdot\mathbf{n}=0$ on $\partial V$. For the magnetic fields described in this paper, this condition is not met as field leaves the stellar surface and either returns back down to this level (closed field) or connects to the source surface above (open, stellar wind bearing, field). \cite{berger84} show that a gauge-invariant measure of helicity is also possible for open magnetic fields. This form of helicity is known as relative helicity and can be  written as
\begin{equation}
H_R = H(\mathbf{B}_V, \mathbf{B}_{V_-}, \mathbf{B}_{V_+})  -  H(\mathbf{B}^{\rm ref}_V, \mathbf{B}_{V_-}, \mathbf{B}_{V_+}),
\end{equation}
where
\begin{equation}
H(\mathbf{B}_{V_1},\dots,\mathbf{B}_{V_n}) = \sum_{i=1}^n\int_{V_i}\mathbf{B}_{V_i}\cdot\mathbf{A}_{V_i}\,{\rm d}V.
\end{equation}

Here $V$ is the volume between the stellar surface and the source surface and $V_-$ and $V_+$ represent volumes below and above $V$ respectively. The magnetic fields in these two volumes match the field of $V$ at its boundaries and close to allow for helicity to be a topological invariant. \cite{berger84} show that the magnetic fields in $V_-$ and $V_+$ can be ignored and that helicity only requires knowledge of the field in $V$. The required magnetic fields are $\mathbf{B}_V$, the coronal magnetic field found from equation (\ref{globalB}), and $\mathbf{B}_V^{\rm ref}$, a reference magnetic field with the same boundary conditions as $\mathbf{B}_V$.  The reference field is normally taken to be a potential magnetic field for simplicity. Relative helicity can then be thought of as a quantitative measure of how `different topologically' $\mathbf{B}_V$ is from $\mathbf{B}_V^{\rm ref}$.

\begin{figure}
\includegraphics[width=\linewidth]{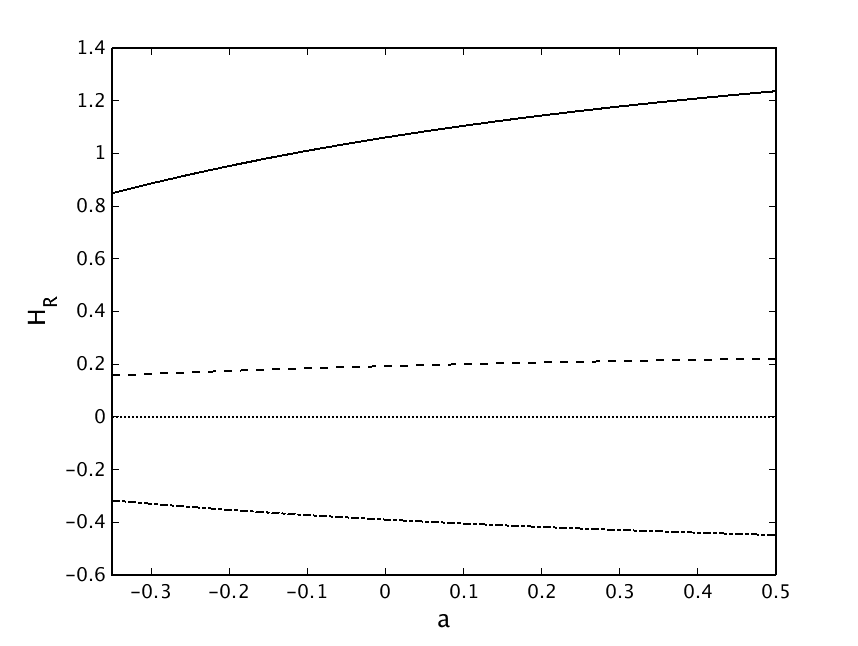}
\caption{ The variation of relative helicity $H_R$ as a function of $a$ for different values of $\alpha$.  Key: solid, $\alpha=0.5$; dashed, $\alpha=0.1$; dotted, $\alpha=0$; dot-dashed, $\alpha=-0.2$. Values on the axes are non-dimensional as specified in equation (\ref{nondim_p}).}
\label{hr_fig}
\end{figure}

If $\mathbf{B}_{V_-}$ and $\mathbf{B}_{V_+}$ are chosen to be potential, the relative magnetic helicity reduces to
\begin{equation}
H_R = H(\mathbf{B}_V, \mathbf{B}_{V_-}, \mathbf{B}_{V_+}),
\end{equation}
in a spherical geometry. In order to determine an expression for $H_R$, we proceed by decomposing the magnetic field into poloidal $(P)$ and toroidal $(T)$ components
\begin{equation}\label{poltor}
\mathbf{B} = \nabla\times\nabla\times(P\mathbf{r}) + \nabla\times(T\mathbf{r}).
\end{equation}
 The magnetic field in equation (\ref{poltor}) can be written as 
\begin{equation}
\mathbf{B}=-i[\nabla\times(\mathbf{L}P)+\mathbf{L}T].
\end{equation}
The vector potential is found to be
\begin{equation}
\mathbf{A} = -i[\mathbf{L}P+\mathbf{r}T+\nabla\lambda],
\end{equation}
where $\lambda$ is an arbitrary gauge function. The relative helicity can be expressed in the appealing form
\begin{equation}\label{rel_hel}
H_R=-2\int_V\mathbf{L}P\cdot\mathbf{L}T\,{\rm d}V,
\end{equation}
which reflects the fact that helicity arises due to the linkage of the poloidal and toroidal magnetic field. Equation (\ref{rel_hel}) (in a slightly different form) is derived in \cite{berger85} and so we do not repeat the derivation here. \cite{berger85} then goes on to find an expression for the relative helicity of linear force-free fields. We shall now proceed in a similar manner and find an expression for this class of MHS equilibria. Inserting equation (\ref{poltor}) into equation (\ref{current}), we can derive the equations
\begin{equation}
\nabla^2P+\alpha T+F = 0, \quad T-\alpha P = 0,
\end{equation}
which can be reduced to 
\begin{equation}\label{P}
\nabla^2P+\alpha^2P+F = 0.
\end{equation}
Using equation (\ref{poltor}), we can show that
\begin{equation}
\mathbf{L}^2P = \mathbf{r}\cdot\mathbf{B}.
\end{equation}
Applying $\mathbf{L}^2$ to equation (\ref{P}) and using the distributive property of $\mathbf{L}^2$, we obtain
\begin{equation}
\nabla^2(\mathbf{r}\cdot\mathbf{B}) + {\xi}(r)\mathbf{L}^2(\mathbf{r}\cdot\mathbf{B})+ {\alpha}^2(\mathbf{r}\cdot\mathbf{B}) = 0,
\end{equation}
which is just equation (\ref{vec_schrod}). Hence, one can expand $P$ as
\begin{equation}
P =  \sum_{l=1}^{\infty}\sum_{m=-l}^lQ_{lm}Y^m_l(\theta,\phi).
\end{equation}
 Following \cite{berger85}, after integration by parts, equation (\ref{rel_hel}) with suitable substitutions becomes
\begin{equation}\label{rhe}
H_R = 2\alpha\sum_{l=1}^{\infty}\sum_{m=-l}^ll(l+1)\int_{R_{\ast}}^{R_{SS}}|Q_{lm}|^2r^2\,{\rm d}r,
\end{equation}
where $R_{\ast}$ is the stellar radius and $R_{SS}$ is the radius of the source surface. To investigate the dependence of the relative helicity on the free parameters, we determine equation (\ref{rhe}) for the dipolar field with boundary conditions specified in  equation (\ref{bcs}). Figure \ref{hr_fig} displays how $H_R$ varies as a function of $a$ for different values of $\alpha$.

The curves in Figure \ref{hr_fig} can be interpreted as how much the field is `different topologically' from a potential field with the same boundary conditions.  For $\alpha=0$, $H_r=0$, as shown by the dotted line in Figure \ref{hr_fig}. If there is no twist injected into the field, through $\alpha$, then the field is topologically equivalent to a potential field and so $H_R=0$. For $\alpha>0$, the relative helicity increases monotonically with $a$. For $\alpha<0$, $|H_R|$ also increases monotonically but now the sign of the relative helicity is negative. In the range of the parameter space displayed in Figure \ref{hr_fig}, the effect of increasing both $a$ and $\alpha$ is to increase the magnitude of the relative helicity. This result occurs because increasing the free parameters $a$ and $\alpha$ increases the current density which, in turn, moves the field further from a potential state. For much higher values of $\alpha$, resonances can occur where the magnetic helicity and magnetic energy go to infinity \citep{berger85}. These large values, however, are not applicable to our model which is concerned with large-scale stellar fields that are not twisted to such extents \citep[e.g.][]{zhao00}.

\section{Application}

\begin{figure}
\includegraphics[width=\linewidth]{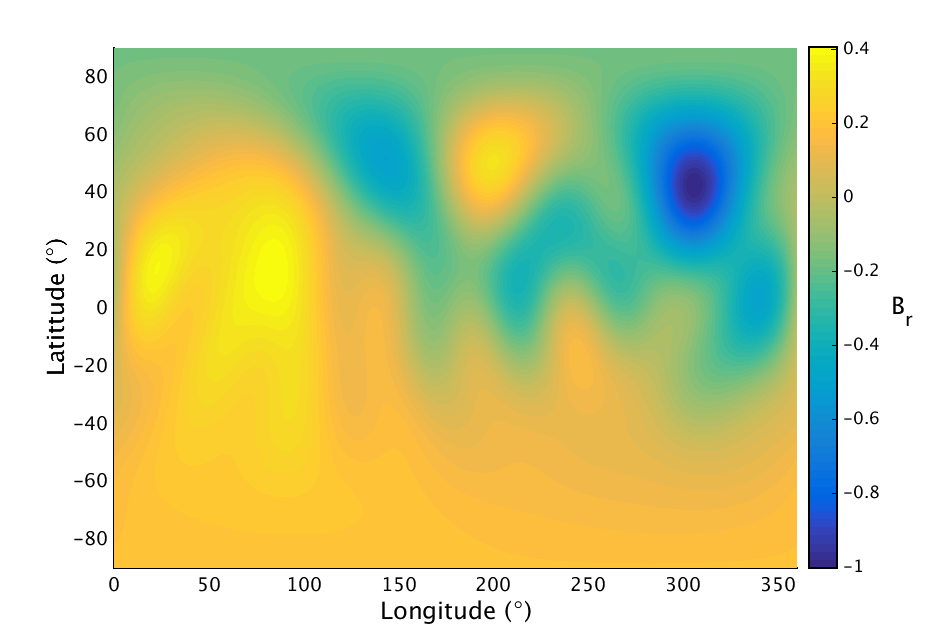}
\caption{Magnetogram of GJ~182. $B_r$ has been normalized with respect to its largest magnitude (355~G).}
\label{br_fig}
\end{figure}

To illustrate the model, we apply it to the young single M-dwarf GJ~182 ($M_{\ast}=0.75\,{\rm M}_{\odot}$, $R_{\ast}=0.82\,{\rm R}_{\odot}$, $P_{\rm rot}=4.35\,{\rm d}$; \cite{donati08}). This star shows evidence for surface differential rotation, with the equatorial regions rotating faster than the poles. The magnetic field is dominantly toroidal with a mostly non-axisymmetric poloidal component \citep{donati08}, consistent with a partially convective interior \citep{gregory12}. As input to the model, we take  GJ~182's magnetogram, the radial component of the magnetic field at the stellar surface. The magnetogram is shown in Figure \ref{br_fig} where the field is normalized with respect to its largest magnitude, 355~G.  This star has been chosen to demonstrate the model because the star's magnetic field differs from the large-scale symmetric dipolar structure of the solar magnetic field. Also,  GJ~182's coronal magnetic field has been modelled by \cite{lang12} using a potential field extrapolation. This previous study allows us to make a comparison and highlight the extra features that appear in our model.

%In the previous section, the expressions for $B_r$ at the stellar surface were bipoles. Hence, the total flux at the surface was zero. This is a condition that must be satisfied in order to find a correct solution. For a physical interpretation of this, consider the solenoidal condition (\ref{divB_nd}). As this holds in $\mathbb{R}^3$, we have
%\[
%0 = \int_V\nabla\cdot\mathbf{B}\,{\rm d}V = \int_{\partial V}\mathbf{B}\cdot\mathbf{n}\,{\rm d}S,
%\]
%by an application of the divergence theorem. Here $V \subset \mathbb{R}^3$ is the volume enclosed by the star and $\partial V$ is its surface. So for a spherical star where the surface flux is determined purely by the radial component of the field, the flux must be zero. If this is not the case, magnetic monoples are present in the interior of the star. Typically, magnetograms are not in flux balance due to observational or interpolational errors. In order to correct the flux balance without changing the topology of the magnetogram, we perform the following procedure:
%\begin{itemize}
%\item Calculate the total flux $F=\int\mathbf{B}\cdot\mathbf{n}\,{\rm d}S$.
%
%\item Divide $F$ by the number of grid cells where $B_r\ne 0$, $n_0$ say.
%
%\item Subtract $F/n_0$ from each cell with non-zero $B_r$.
%\end{itemize}
%Using this procedure, we do not change polarity boundaries where $B_r=0$ and, hence, preserve the topology. Performing this for the magnetogram of GJ~182 results in $F= O(10^{-11})$. 

\begin{figure}
\includegraphics[width=\linewidth]{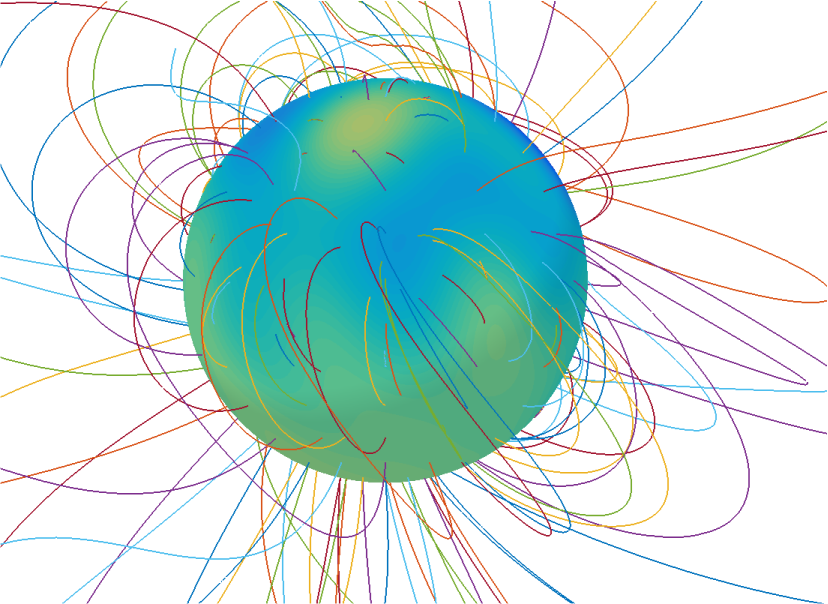}
\caption{A field line extrapolation using $\alpha=0.4$ and $a=0.3$. The contour map shows $B_r$. Sigmoidal field lines are representative of $\mathbf{j}\ne\mathbf{0}$. Field lines are multi-coloured to make them easier to visualize.}
\label{fieldline_fig}
\end{figure}

To complete the model, an outer boundary condition, the source surface, is required.  As mentioned previously, the radial position and form of the source surface are modelling choices.  For simplicity, we take the dipolar radial field
\begin{equation}
B_r(r=4) = \beta Y^0_1(\theta,\phi),
\end{equation}
where $\beta = O(r^{-3})$. With these radial boundary conditions, the magnetic field in the region $1<r<4$ can be calculated using equation (\ref{globalB}). We take $l=8$, which is enough to capture the features of the magnetogram. In the last section, we discussed some of the effects of the free parameters. For applying the model to GJ~182, we choose $\alpha=0.4$ and $a=0.3$. This choice lies in the part of the parameter space where the magnetic pressure is increased, resulting in a pressure deficit due to the magnetic field. A visualization of the coronal field is shown in Figure \ref{fieldline_fig} where several field lines are plotted and the sphere shows the map of $B_r$ given in Figure \ref{br_fig}.  As mentioned above, \cite{lang12} plot a potential field extrapolation of the corona of GJ~182.  The orientation of the star in Figure \ref{fieldline_fig} is chosen to be similar to that shown in \cite{lang12} for ease of comparison.

Although the source surface will help to determine the behaviour of open field lines, its influence on the geometry of lower closed field lines is smaller as such is  determined primarily by the form of the current density. Looking at the field lines between the large positive and negative regions of $B_r$, the arcades are sheared. That is, the low field lines at the polarity inversion line are at a different angle to those above. As mentioned before, this effect is due to a non-zero current. The twist of the field due to the linear force-free term can also be seen in the sigmoidal shape of the field lines. This shape is most clearly visualized in Figure \ref{fieldline_fig} by the higher field lines of the sheared arcade. Reducing the magnitude of $\alpha$ would result in field lines becoming straighter, i.e. losing their sigmoidal shape. In the potential extrapolation of \cite{lang12}, the arcade field lines do not possess a sigmoidal shape as $\alpha=0$ in their case. The expansion of the field lines can be controlled by $a$, as described previously.

\begin{figure}
\includegraphics[width=\linewidth]{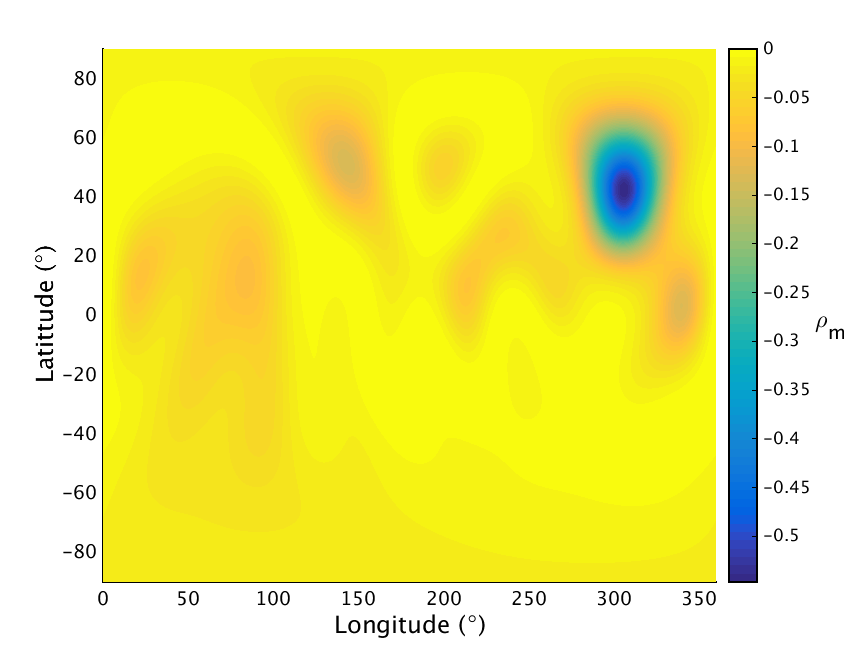}
\caption{The density perturbation, $\rho_m = r^2[\xi'(r)(\mathbf{r}\cdot\mathbf{B})^2/2+r\xi(r)\mathbf{B}\cdot\nabla(\mathbf{r}\cdot\mathbf{B})]$, at the stellar surface.  Values on the colour scale are non-dimensional as specified in equation (\ref{nondim_p}).}
\label{density_fig}
\end{figure}

To see the effects of the coronal magnetic field on the background atmosphere, we have constructed maps of the density and pressure perturbations at the stellar surface. Figure \ref{density_fig} shows the density perturbation, the second term on the RHS of equation (\ref{density}).

\begin{figure}
\includegraphics[width=\linewidth]{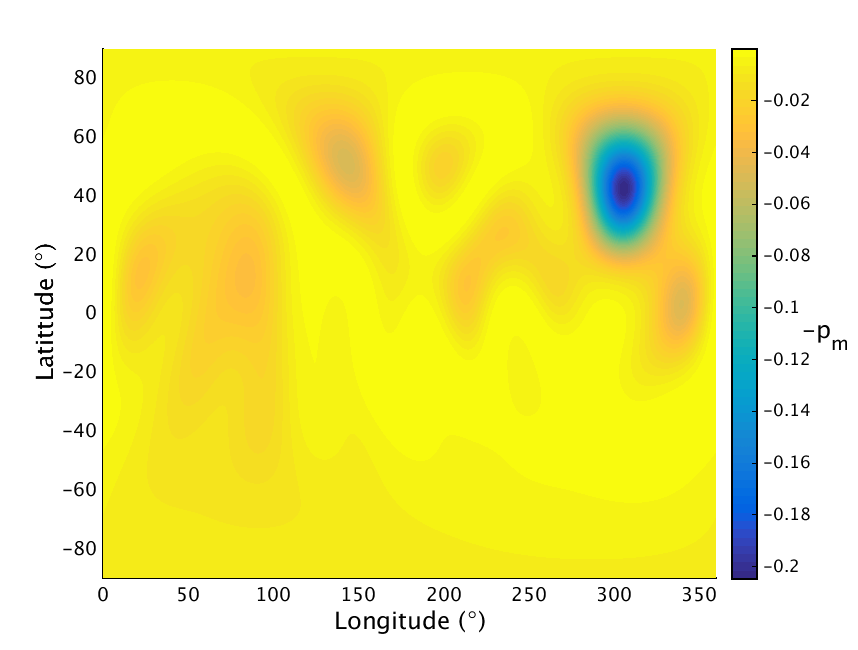}
\caption{The pressure perturbation, $-p_m = -\xi(r)(\mathbf{r}\cdot\mathbf{B})^2/2$, at the stellar surface. The values are negative as we are considering the perturbation to the hydrostatic pressure, shown in equation (\ref{pressure}).  Values on the colour scale are non-dimensional as specified in equation (\ref{nondim_p}).}
\label{pressure_fig}
\end{figure}

In the magnetogram for GJ~182 (Figure \ref{br_fig}), the radial field in the northern hemisphere contains concentrations that alternate in sign. Beginning with a positive concentration, this changes to negative with increasing longitude. Moving further around the star in longitude, the positive to negative pattern repeats.  This pattern is also revealed in the density perturbation map. From equation (\ref{density}) the density perturbation is proportional to $B_r^2$. Hence, the stronger the magnetic polarity, the greater the density perturbation. The pressure perturbation map, shown in Figure \ref{pressure_fig}, is very similar to the density perturbation map. Again, this quantity, the second term on the RHS of equation (\ref{pressure}), is proportional to $B_r^2$. The only significant difference between the two maps is the magnitude of the perturbations.

\section{Summary}

In this paper we have presented, analyzed and demonstrated a MHS model for stellar coronae. The model extends the popular potential field model to allow for the magnetic field to contain current and couple to the background density and pressure. The main assumption of the model is the choice for the current density, which is split into a linear force-free part and another part that links the field to the background atmosphere. The model, given by equation (\ref{current}), has its particular form so that Amp\`{e}re's law (\ref{ampere}) is linear and analytical progress is possible. The MHS equations then boil down to solving Bessel's equation subject to a model choice for $\xi(r)$. In this paper, we have chosen a form that allows for analytical progress. Other choices that allow for analytical solutions are also possible \citep[e.g.][]{neukirch95}. Of course, more complicated expressions for $\xi(r)$ can be specified but will require a numerical treatment \citep[e.g.][]{dmac13}. A closed-form expression for the coronal magnetic field is found and the associated perturbations to the background hydrostatic pressure and density are also determined.

We present an analysis of the effects of the two free model parameters. In terms of the geometry of the field, the parameter $\alpha$ increases the shear/twist in the magnetic field. Increasing the magnitude of $\alpha$ creates field lines with a sigmoidal shape. The parameter $a$ is related to the inflation of magnetic field lines through the magnetic pressure. The size of magnetic arcades can change depending on the choice of $a$. The topology of the coronal field is studied by considering the relative magnetic helicity. This quantity gives a  measure of how different, topologically, the coronal field is compared to a potential field with the same boundary conditions (the choice of reference field in this paper). An expression for the relative helicity, for the class of MHS equilibria presented here, is determined. Increasing the magnitude of $\alpha$ also increases the relative helicity. This increase occurs because the current, and hence twist, increases in the magnetic field. Increasing the magnitude of $a$ increases the relative helicity.

Finally, a simple demonstration of the model is given for the M0.5 dwarf star GJ~182. Field lines are plotted showing the non-potential features of the model. The density and pressure perturbations at the stellar surface are also displayed and are proportional to $B_r^2$ so that shapes on the countour plots can be matched to corresponding ones on the magnetogram.

The model presented in this paper is suitable for determining the large-scale features of a star's corona. The model also has the advantage of being simple, fast to implement and, unlike full MHD models, is not computationally expensive to implement. The effort involved is only slightly more than a potential field model (which is a special case of the MHS model). By a suitable choice of $\xi(r)$ and the free parameters $\alpha$ and $a$, one can investigate stellar coronae with different geometries, topologies, pressure and density distributions and current density profiles. In future work, we will apply our model to stellar observations and attempt to constrain the free parameters using the information provided by the observations, such as the toroidal component of the magnetic field.

\section*{Acknowledgements}
SGG acknowledges support from the Science \& Technology Facilities Council (STFC) via an Ernest Rutherford Fellowship [ST/J003255/1]. TN would like to acknowledge support from the STFC Consolidated Grant ST/K000950/1. The authors would like to thank J. Morin for helpful advice on the paper.

\appendix
%
%\section{Forms of $\xi$}\label{app_xi}
%In order to find analytical expressions for equations (\ref{u1}) and (\ref{u2}), a particular form for $\xi(r)$ must be chosen. This can be achieved by noticing that equation (\ref{g}), reproduced below for convenience,
%\begin{equation}\label{ag}
%\left[\frac{{\rm d}^2 }{{\rm d}r^2} -l(l+1)\left(\frac{1}{r^2}-\xi(r)\right) + {\alpha}^2\right]g_l^{(j)}=0,
%\end{equation} 
%is of a similar form to a 1D Schr\"{o}dinger equation \citep[e.g.][]{rae02}. Here, $\xi(r)$ plays the role of a potential function.  This analogy gives us a clue as to what possible forms $\xi(r)$ could take. Two possible choices

\section{Spherical components}\label{appendix}
The spherical components of the coronal magnetic field are
\begin{equation}
B_r = \sum_{l=1}^{\infty}\sum_{m=-l}^{l}\sum_{j=1}^2A^{(j)}_{lm}\frac{l(l+1)}{r}u_l^{(j)}(r)Y^m_l(\theta,\phi),
\end{equation}

\begin{eqnarray}
B_{\theta} &=& \sum_{l=1}^{\infty}\sum_{m=-l}^{l}\sum_{j=1}^2A^{(j)}_{lm}\left[\frac{\alpha}{\sin\theta}u_l^{(j)}(r)\frac{\partial}{\partial\phi}Y^m_l(\theta,\phi)  \right.  \nonumber \\ 
 &+&\left.\frac{1}{r}\frac{{\rm d}}{{\rm d}r}(ru_l^{(j)}(r))\frac{\partial}{\partial\theta}Y^m_l(\theta,\phi)\right], 
\end{eqnarray}
and
\begin{eqnarray}
B_{\phi}&=&\sum_{l=1}^{\infty}\sum_{m=-l}^{l}\sum_{j=1}^2A^{(j)}_{lm}\left[-\alpha u_l^{(j)}(r)\frac{\partial}{\partial \theta}Y^m_l(\theta,\phi) \right. \nonumber \\
&+& \left.\frac{1}{r\sin\theta}\frac{{\rm d}}{{\rm d}r}(ru_l^{(j)}(r))\frac{\partial}{\partial\phi}Y^m_l(\theta,\phi)\right]. 
\end{eqnarray}
All terms are defined as in the main body of the paper.

\end{document}